# Designing a Novel Method for Personalizing Recommendations to Decrease Plastic Pollution

Seung Ah Choi1, Monta Vista High School, Cupertino, California

## I. INTRODUCTION

Individuals of first world countries that already have streamlined waste management systems, who do not wish to contribute to the plastic pollutants that harm oceans and wildlife health, but also do not have enough belief that individual action can make a significant change, thus continuing to use single-use plastics out of convenience, need a mobile application that encourages these individuals to work towards zero-waste through small actions (ex.using a mug at Starbucks or using a reusable fork at school) by visibly quantifying all the small actions (through possibly a point system) that happen in a day regarding how much plastic was saved. The first analysis is to compare whether the call for governmental/individual action, in general, is higher in first or third world countries. After analyzing multiple Reddit threads that correspond to the topic, turns out that the count of words that call for governmental/individual action is greater in response to plastic pollution in third-world countries rather than in first-world countries and the count of words that are apathetic towards governmental/individual action is greater in response to plastic pollution in first-world countries than in third-world countries.

The second analysis is to compare whether zero waste action is more prominent in first world countries or third world countries. After analyzing multiple Reddit threads that correspond to the topic, turns out that the count of supporting/positive words will be greater in response to actions for zero waste than to actions against zero waste and the count of opposing/negative words will be greater in response to actions against zero waste than to actions for zero waste.

## II. ABSTRACT

According to Our World in Data, Third World countries (i.e. India) tend to have a higher share of plastic waste that is inadequately managed while First World countries (i.e. The US) have higher plastic waste generation per person. This difference in the characteristics of plastic pollution depending on the country's standing results in varying optimal recommendations for users depending on which country they live in. Through Big Text and OSOME meme analysis, I constructed a list with optimal recommendations for First World and Third World countries. Based on the list, I designed a User Interface (UI) with Google Apps Scripts that provide personalized recommendations based on the country's standing and user's preferred difficulty and reassessed the code based on the six qualities of code. The User Interfaces' purpose is to aid people who wish to help solve plastic pollution by offering a set of personalized tasks for each user and keeping their progress accountable through a point tracking system. With a significant number of users, the application could eventually contribute to solving plastic pollution.

## Big Text Data Analysis

The first analysis is to compare whether the call for governmental/individual action, in general, is higher in first or third world countries. After analyzing multiple Reddit threads that correspond to the topic, turns out that the count of words that call for governmental/individual action is greater in response to plastic pollution in third-world countries rather than in first-world countries and the count of words that are apathetic towards governmental/individual action is greater in response to plastic pollution in first-world countries than in third-world countries.

The second analysis is to compare whether zero waste action is more prominent in first world countries or third world countries. After analyzing multiple Reddit threads that correspond to the topic, turns out that the count of supporting/positive words will be greater in response to actions for zero waste than to actions against zero waste and the count of opposing/negative words will be greater in response to actions against zero waste than to actions for zero waste.

## First Big Text Data Analysis

**Hypothesis**: The count of words that call for governmental/individual action will be greater in response to plastic pollution in third-world countries rather than in first-world countries and the count of words that are apathetic towards governmental/individual action will be greater in response to plastic pollution in first-world countries than in third-world countries.

**Null Hypothesis**: The count of words that call for governmental/individual action and the count of words that are apathetic towards governmental/individual action will be approximately the same in response to plastic pollution in first-world countries and in third-world countries.

*Figure 1. Word Cloud for Plastic Pollution in Third-World Countries and in First-World countries*

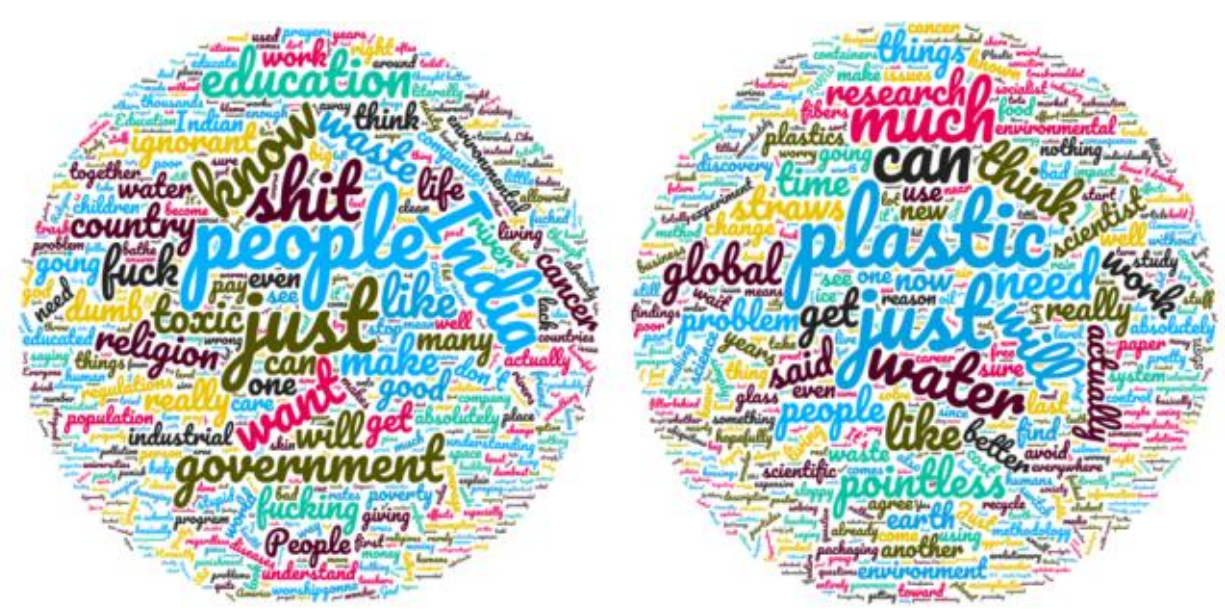

Both word clouds were generated by a text that was a compilation of comments from a Reddit thread which was about plastic pollution in each first-world and third-world country. The body of the text was both around 3.2k words, and I chose the comments from most upvoted in order so that there was no bias involved in my selection of the comments. I chose the particular Reddit threads because it was the most upvoted/discussed topic amongst plastic pollution in third-world and first-world countries. At first glance, the most frequently used words in both word clouds (or the words that are the largest size) are general topics like ‘people,’ ‘India,’ ‘plastic,’ ‘water’ etc. Therefore I will focus more on the words that are related to the alternative and null hypothesis that I have generated. I believe that the word clouds support my alternative hypothesis more than the null hypothesis.  The first Word Cloud that depicts people’s responses to plastic pollution in third-world countries (in this case India, as it is one of the most used words in the word cloud) mainly points to how people think that the main problem regarding plastic pollution in third-world countries is that people are uneducated as shown by the repetitive use of the words ‘ignorant’ and ‘dumb’. Moreover, as depicted by the repetitive use of the words ‘government’ and ‘education,’ people strongly advocate for governmental action in India to educate the ‘people’ (another word that shows up frequently) the importance of zero waste and reduction of plastic pollution in rivers. This also implies that people generally think that individuals themselves must also take the initiative to learn and get information about how bad plastic pollution is in their country. On the contrary, the second Word Cloud that depicts people’s responses to plastic pollution in first-world countries includes words like ‘pointless’ which is apathetic towards an individual or even governmental change. In this context, the words ‘much’ and ‘can’ also seem to support the idea that people in first-world countries believe that there is not much they can do individually or even as a country. In addition, as depicted by the frequent use of ‘global’ and ‘research,’ the people of the first-world country called for global action and more scientific studies on how to decrease the amount of plastic waste.

**Table 1:** In order to get an accurate count from each word cloud which is case sensitive, for each word that I chose, I looked for both capitalized and non-capitalized words.

| | Plastic pollution in third-world countries | Plastic pollution in first-world countries |
|---|---|---|
| Call for individual/governmental action (government, education, ignorant, dumb) | 39 | 3 |
| Apathetic towards governmental/individual action (pointless, irrelevant, impossible) | 2 | 11 |

**Table 2. Chi-Squared Test:**

| Results | | | | | | |
|---|---|---|---|---|---|---|
| | Plastic pollution in third-world countries | Plastic pollution in first-world countries | | | | ***Row Totals*** |
| Call for individual/governmental action | 39 (31.31) [1.89] | 3 (10.69) [5.53] | | | | 42 |
| Apathetic towards governmental/individual action | 2 (9.69) [6.10] | 11 (3.31) [17.88] | | | | 13 |
| | | | | | | |
| | | | | | | |
| | | | | | | |
| ***Column Totals*** | 41 | 14 | | | | 55 (Grand Total) |

The chi-square statistic is 31.4007. The *p*-value is < .00001. The result is significant at $p < .05$.

This means that the alternative hypothesis, that the count of words that call for governmental/individual action will be greater in response to plastic pollution in third-world countries rather than in first-world countries and the count of words that are apathetic towards governmental/individual action will be greater in response to plastic pollution in first-world countries than in third-world countries, is supported.

**Conclusion**: Now that I have two kinds of analysis (word clouds and chi-squared), I feel that the alternative hypothesis is supported. From the chi-squared test, the p-value is <.00001, and is significant at p<.05. From the word cloud, I saw that the majority of the words that call for governmental and individual action in the word cloud for third-world countries, and I saw the majority of the words that were apathetic towards governmental and individual action in the word cloud for first-world countries. I believe that both the word cloud and the chi-squared test confirms the alternative conclusion.

## Second Big Text Data Analysis

**Hypothesis**: The count of supporting/positive words will be greater in response to actions for zero waste than to actions against zero waste and the count of opposing/negative words will be greater in response to actions against zero waste than to actions for zero waste.

**Null Hypothesis**: The count of supporting/positive words and the count of opposing/negative words will be approximately the same in response to actions for zero waste and to actions against zero waste.

*Figure 2. World Cloud Comparison between the Actions for Zero Waste (Positive) and the Actions against Zero Waste (Negative)*

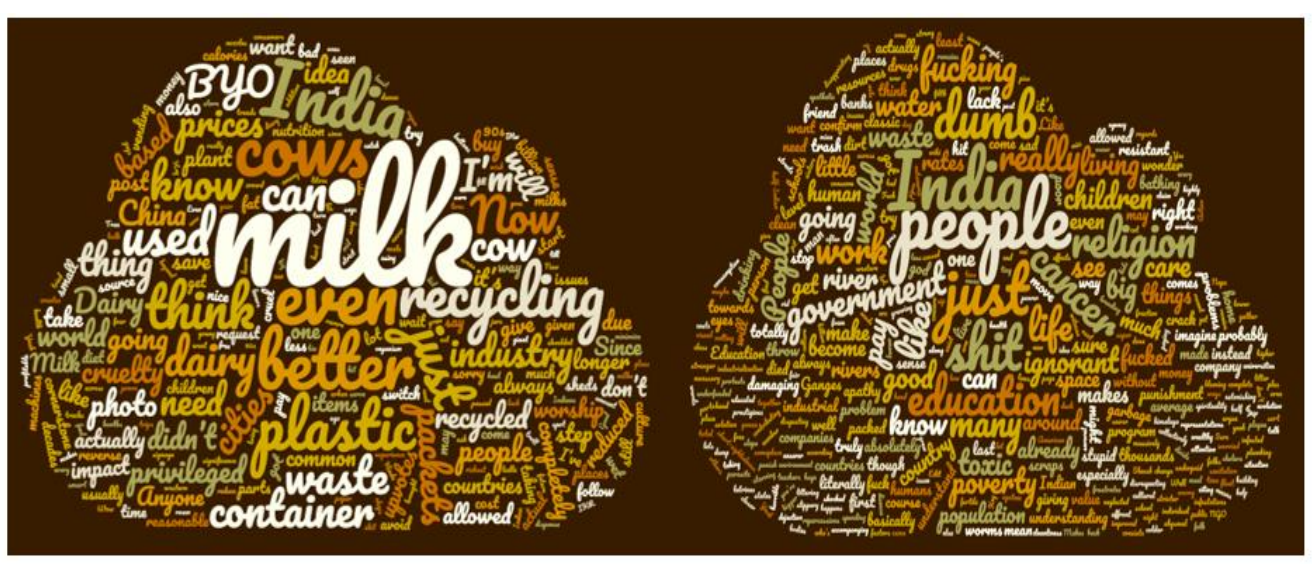

**Table 3**

| | Actions for zero waste (Positive) in India | Actions against zero waste (Negative) in India |
|---|---|---|
| Supporting/Positive Messages (better, good, awesome, nice) | 15 | 4 |
| Opposing/Negative Messages (dumb, sh*t, f*cking, toxic, f*cked, awful) | 2 | 27 |

| | Actions for zero waste (Positive) | Actions against zero waste (Negative) | *Marginal Row Totals* |
|---|---|---|---|
| **Supporting/Positive Messages** | 15 (6.73) [10.17] | 4 (12.27) [5.57] | 19 |
| **Opposing/Negative Messages** | 2 (10.27) [6.66] | 27 (18.73) [3.65] | 29 |
| ***Marginal Column Totals*** | 17 | 31 | 48 (Grand Total) |

The chi-square statistic is 26.0531. The *p*-value is < 0.00001. Significant at $p < .05$.

The chi-square statistic with Yates correction is 22.9984. The *p*-value is < 0.00001. Significant at $p < .05$.

From the chi-squared test, the p-value is <.00001 and is significant at p<.05, which means that the alternative hypothesis from above is supported.

## Osome Meme Analysis

I discussed what other topics are discussed when #plasticpollution is discussed on social media. I used an Osome tool to draw the following conclusions.

Hashtags/Topic: #zerowaste, #cancer, #plasticpollution, #plastic (impacts society)

**Null Hypothesis**: There will not be much of a correlation between #zerowaste or #cancer and #plasticpollution (plastic)

**Alternative Hypothesis**: There will be a strong correlation between #zerowaste or #cancer and #plasticpollution as these two topics have been discussed together frequently.

*Figure 3. Co-Occurrence of tags (networks) #zerowaste, #plasticpollution*

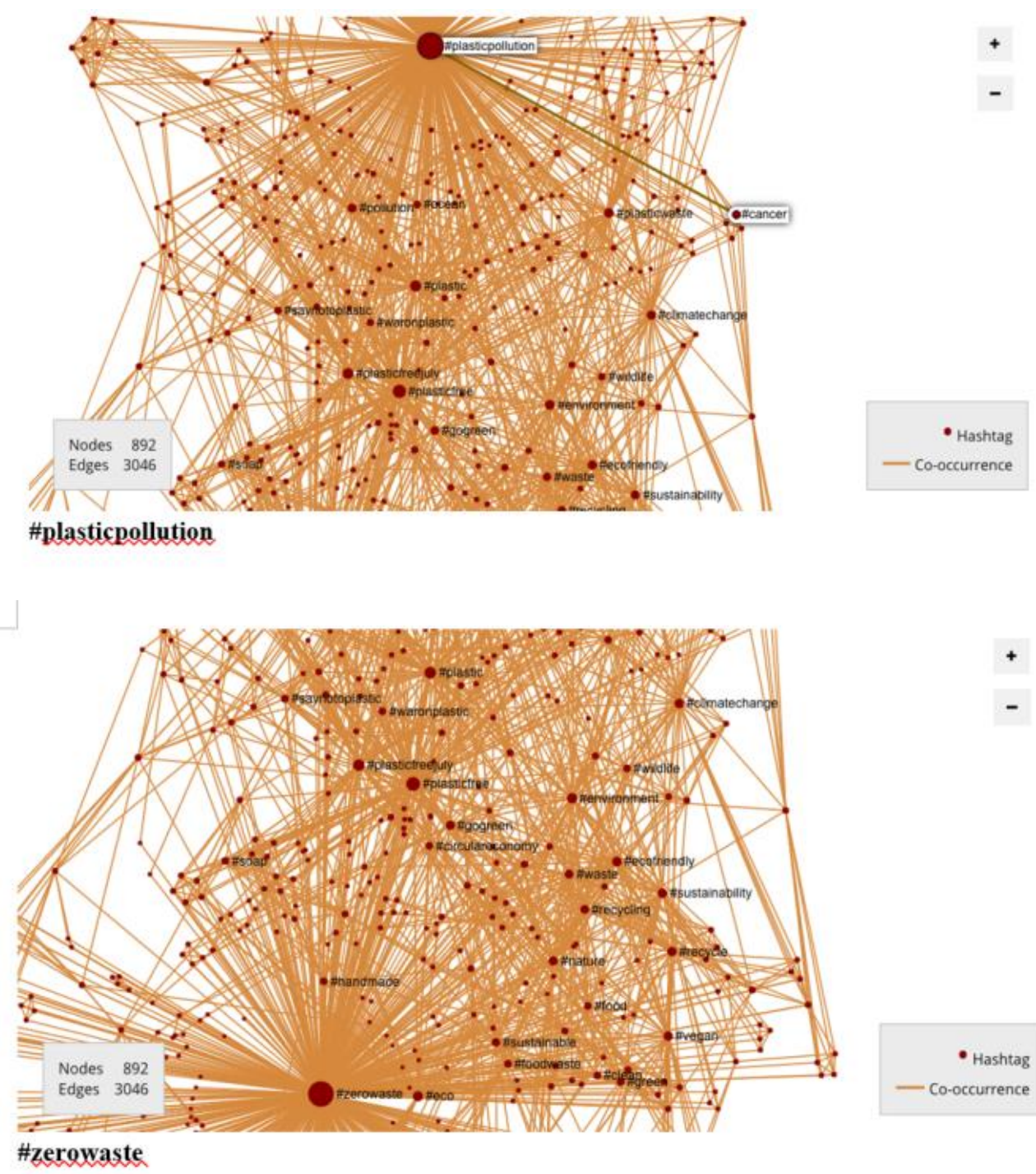

*Figure 4. Time plot*

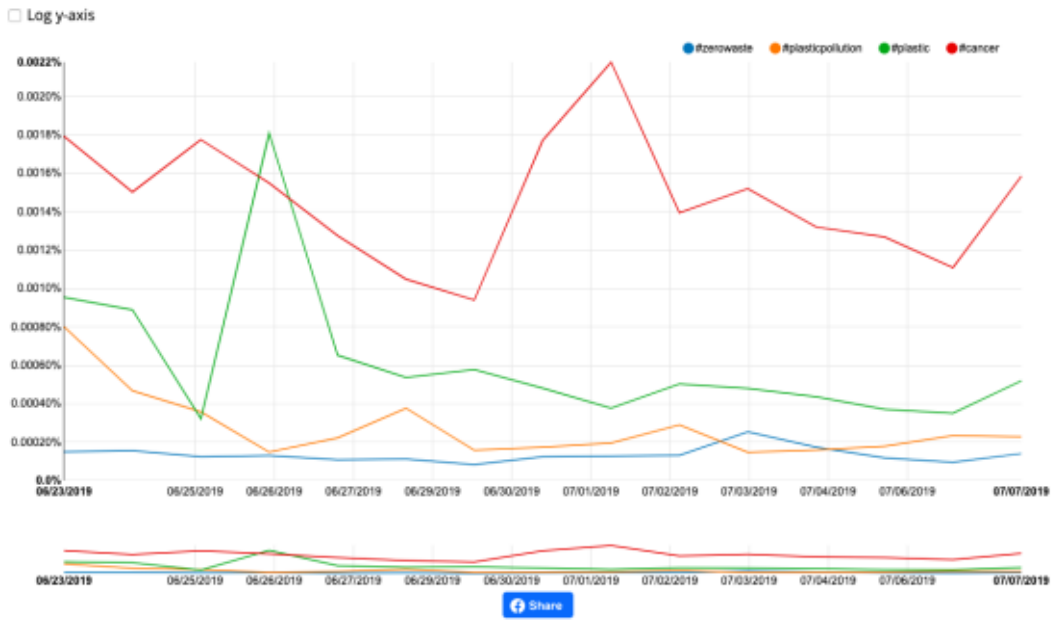

*Figure 5. The same graph as Fig 4, but the natural log version is underneath*

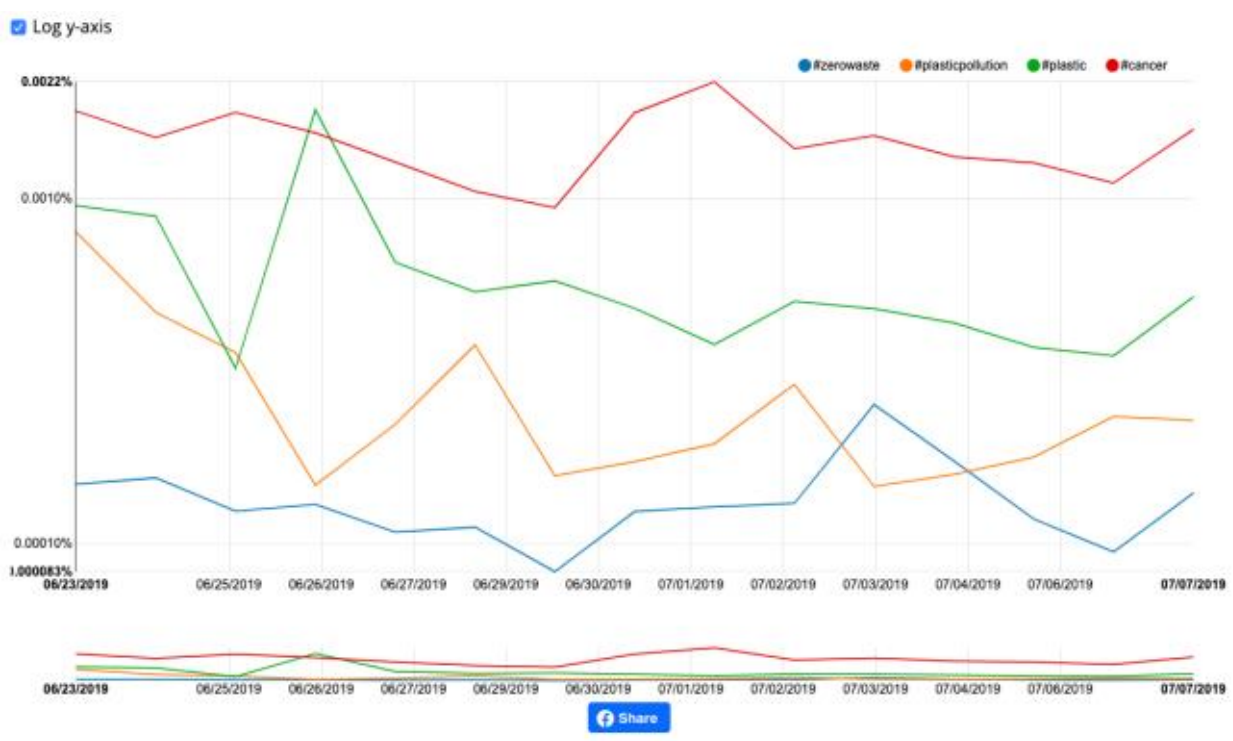

**Conclusion:** From the network graphs for #zerowaste and #plasticpollution, the hashtag that showed the largest incidents of co-occurrence was #cancer. Since the larger the circle in the network graph indicates that the particular hashtag is discussed more by people alongside the given hashtags, this means that #cancer is frequently discussed when people talk about #zerowaste and #plasticpollution. This makes sense to me because health is one of the most significant discussion topics for people. Other hashtags related to #plasticpollution was #ocean since the oceans and rivers are largely impacted by plastic pollution. Environmentally friendly advocation hashtags were also found like #ecofriendly, #waronplastic, and #gogreen. Hashtags related to zero waste were mostly advocating for various methods to go zero waste like #recycling, #handmade, #vegan, and #circulareconomy.

One hashtag that I thought was particularly significant was #plasticfreejuly. When I did research to find the meaning of this hashtag apparently the national zero waste day takes place every year on July 3rd. Based on this information, I created time plot graphs from June 23rd, 2019 to July 7th, 2019, the time period that I thought these hashtags would be discussed most frequently amongst netizens. Turns out the usage of zero waste did significantly increase (around 50%) on July 3rd in particular. The number of users for the hashtag #cancer also increased right before July 3rd significantly, which I believe was a method for internet users to bring about awareness for #plasticpollution and the importance of zero waste before the national zero waste day. The usage of #plastic suddenly increased on June 26th, and while I do not know exactly why, I believe there must have been some law passed about plastic, or some trend regarding recycling or remaking plastic was trending on social media. It may have also been influenced by some viral picture related to plastic, however, it is not possible to know if plastic was used in a negative or positive connotation. Regardless, my guess is that it was used positively because the use of #plasticpollution simultaneously decreased during that time, showing the opposite of co-occurrence. In general, the #cancer was used more than #plastic because people usually care a lot about their health, and these two hashtags were used more on average than #plasticpollution and #zerowaste because it is less specific to a certain topic and can be used on a general basis.

## In-Sheet Calculation Analysis

I calculated the average values for the share of plastic waste that is inadequately managed around the world which turned out to be 34.43% on average and the average value for plastic waste generation per person which turned out to be 0.18ton per year.

**Objective**: Use Google Apps Script in Google Sheets to calculate the average, minimum, maximum, median, and standard deviation value of the share of plastic waste that is inadequately managed in each country and plastic waste generation per person by country. I have calculated the average, minimum, median, maximum, and standard deviation of the two different data sets that I have found each with 186 data points. I have included the calculation formulas that I have used in Table 3.

**Table 3**

| <Share of Plastic Waste That is Inadequately Managed> | | | | <Plastic Waste Generation per Person> | |
|---|---|---|---|---|---|
| Albania | 45 | Share of Plastic Waste That is Inadequately Managed | | Albania | 0.069 |
| Algeria | 58 | AVERAGE(C2:C187) | 34.42702703 | Algeria | 0.144 |
| Angola | 71 | MIN(C2:C187) | 0 | Angola | 0.062 |
| Anguilla | 2 | MEDIAN(C2:C187) | 23 | Anguilla | 0.252 |
| Antigua and Barbuda | 6 | MAX(C2:C187) | 88 | Antigua and Barbuda | 0.66 |
| Argentina | 12 | STDEV(C2:C187) | 34.10591438 | Argentina | 0.183 |
| Aruba | 1 | | | Aruba | 0.252 |
| Australia | 0 | Plastic Waste Generation per Person | | Australia | 0.112 |
| Bahamas | 1 | AVERAGE(I2:I187) | 0.180227027 | Bahamas | 0.39 |
| Bahrain | 10 | MIN(I2:I187) | 0.01 | Bahrain | 0.132 |
| Bangladesh | 87 | MEDIAN(I2:I187) | 0.144 | Bangladesh | 0.034 |
| Barbados | 4 | MAX(I2:I187) | 0.686 | Barbados | 0.57 |
| Belgium | 0 | STDEV(I2:I187) | 0.1232094908 | Belgium | 0.08 |
| Belize | 29 | | | Belize | 0.172 |
| Benin | 83 | | | Benin | 0.043 |
| Bermuda | 0 | | | Bermuda | 0.252 |
| Bosnia and Herzegovina | 40 | | | Bosnia and Herzegovina | 0.144 |
| Brazil | 9 | | | Brazil | 0.165 |
| British Virgin Islands | 0 | | | British Virgin Islands | 0.252 |
| Brunei | 1 | | | Brunei | 0.026 |
| Bulgaria | 31 | | | Bulgaria | 0.154 |
| Cambodia | 87 | | | Cambodia | 0.066 |
| Cameroon | 81 | | | Cameroon | 0.046 |
| Canada | 0 | | | Canada | 0.093 |
| Cape Verde | 74 | | | Cape Verde | 0.065 |
| Cayman Islands | 0 | | | Cayman Islands | 0.252 |

*Table 4. The macro (in google apps script) that I recorded when I was making the formulas*

```
function allFormulas() {
  var spreadsheet = SpreadsheetApp.getActive();
  spreadsheet.getRange('F4').activate();
  spreadsheet.getCurrentCell().setFormula('=AVERAGE(C2:C187)');
  spreadsheet.getRange('F5').activate();
  spreadsheet.getCurrentCell().setFormula('=MIN(C2:C187)');
  spreadsheet.getRange('F6').activate();
  spreadsheet.getCurrentCell().setFormula('=MEDIAN(C2:C187)');
  spreadsheet.getRange('F7').activate();
  spreadsheet.getCurrentCell().setFormula('=MAX(C2:C187)');
  spreadsheet.getRange('F8').activate();
  spreadsheet.getCurrentCell().setFormula('=STDEV(C2:C187)');
  spreadsheet.getRange('F11').activate();
  spreadsheet.getCurrentCell().setFormula('=AVERAGE(I2:I187)');
  spreadsheet.getRange('F12').activate();
  spreadsheet.getCurrentCell().setFormula('=MIN(I2:I187)');
  spreadsheet.getRange('F13').activate();
  spreadsheet.getCurrentCell().setFormula('=MEDIAN(I2:I187)');
  spreadsheet.getRange('F14').activate();
  spreadsheet.getCurrentCell().setFormula('=MAX(I2:I187)');
  spreadsheet.getRange('F15').activate();
  spreadsheet.getCurrentCell().setFormula('=STDEV(I2:I187)');
  spreadsheet.getRange('F16').activate();
};
```

**Conclusion**: The formulas I used in Google Apps Script made it significantly easier for me to calculate the average, minimum, median, maximum, and standard deviation of 186 data points because it allows selecting a group or range of the data sets that I want to explore. If I had to input each data point in a calculator and used the known standard mathematical formulas, the process would have taken much longer.

## Algorithm Overview

The program's purpose is to help people decrease plastic pollution by recommending specific tasks like avoiding using plastic (if their country has high plastic waste generation) or pushing for better waste management systems (if their country has high plastic waste mismanagement). My video explains the significance of optimizing recommendations based on their country and describes the following process. Once the user inputs their country, the code determines the country's standing. Then, the user is asked if they want recommendations. If they answer NO, the app terminates with a thank-you message. If they answer YES, the program asks the user for the preferred difficulty of recommendations. Based on country standing and preferred difficulty, personalized recommendations are called. Once tasks are completed, a specific number of points based on task difficulty is awarded. Finally, the user receives an encouraging message along with the total points they gained.

My peers who test-ran my rough application thought it lacked user interactivity because my code mainly incorporated constants. Upon receiving this feedback, I independently fixed the problem. First, I included method getCurrentCell() to assign variable userAnswer to whichever cell the user selects. This way, the stored value will alter real-time based on the user's choice. However, I soon realized when I run the following methods the currently active cell would change, storing a different value for userAnswer. Thus, I decided to copy the user's initially selected value to a constant range, so the variable stays constant throughout running the code. Second, whenever I rerun the copyOverforPoints() method, the new set of recommendations would paste over the old set because variable targetRow (indicating the row to paste in the recommendations in the 'Point Tracker' Spreadsheet) is constant. I wanted the variable to change based on the user's decisions, but couldn't find a way to save values from a previous run since data resets every run. Therefore, I added pseudocode, which has guidelines on how to change targetRow necessarily. Although directly changing the code may be inefficient and time-consuming, I concluded it's the best way to preserve user interactivity while accumulating data.

[Refer to the code below for the circle] My partner and I worked independently and later combined our methods in proper sequence so the code works together. The following methods can run individually. determineCountryStanding() categorizes the searched country into First-World, Third-World, or Neither. This method has if-else if blocks that contain if-statements that use inequalities based on percentValue (country's percent mismanaged plastic waste) which is searched by my partner's getCountryInfoByName() method. Then, askUserAbtRecommendations() asks if the user wants recommendations to help solve plastic pollution. If-else blocks of this method ensure different outputs based on user response making the code polymorphic. The response NO terminates the program with a thank you message and YES asks for the user's preferred difficulty (HARD, MEDIUM, or EASY). If answered YES, activateGetRec() can correctly run since it can store the user's preferred difficulty (determined by askUserAbtRecommendation()) to variable userAnswer. activateGetRec() calls getRecommendations(userAnswer) which outputs personalized recommendations. This method contains if-statements and boolean values which ensures only the recommendations that match both countryStanding (output of determineCountryStanding()) and userAnswer is copied over from 'Recommendations Data Set' Spreadsheet. For-loops of this method make sure to check every row of the data set by iteratively increasing foundRow and prevent overlaps of pasted recommendations by incrementally increasing targetRow.

[Refer to the code below for the rectangle] My method pointSystem() implements my abstraction. My abstraction is a 2D array. pointSystem() uses logic through if-else if blocks which are divided based on the difficulty levels of the recommendations. This is indicated by the first column of the array (column C in the pointTrackerSpreadsheet). The second column (column D in the pointTrackerSpreadsheet) determines if the task is completed by checking if the user inputted an O. If both criteria are met, point values are outputted accordingly: ten points for HARD levels, five points for MEDIUM, and one point for EASY. pointSystem() uses math when foundRow value incrementally increases by one in the for-loop which causes the program to move onto the next row of the matrix. Math also shows when it calculates the total points in sourceRange using the ‘SUM’ formula Google Apps Script provides. This abstraction makes code more efficient by helping manage the complexity of using two different variables at the same time. Using a 2D array allows me to store both difficulty level and user input. Thus, I can print out corresponding point values dynamically using a for loop, by simply changing the index numbers, making the final total calculation easier without creating many different variables.

```
function determineCountryStanding() {

  var getCountryInfoSpreadsheet
     = SpreadsheetApp.getActiveSpreadsheet().getSheetByName('Get Country Info UI');
  var getCountryInfoRange
     = getCountryInfoSpreadsheet.getRange('F17');
  var percentValue
     = getCountryInfoRange.getValues();
  var getRecSpreadsheet
     = SpreadsheetApp.getActiveSpreadsheet().getSheetByName('Get Recommendations UI');

  var resultsMsg = "The country that you searched for would be considered: "
  var resultsMsgRange = getRecSpreadsheet.getRange('A3:C3').activate();
  getRecSpreadsheet.getRange('A3:C3').mergeAcross();
  resultsMsgRange.setValue(resultsMsg).setFontWeight('bold').setBackground('#9fc5e8');

  var getRecRange1= getRecSpreadsheet.getRange('A4:C4');
      getRecSpreadsheet.getRange('A4:C4').mergeAcross();
  var getRecRange2= getRecSpreadsheet.getRange('A5:C5');
      getRecSpreadsheet.getRange('A5:C5').mergeAcross();
  var getRecRange3= getRecSpreadsheet.getRange('A6:C6');
      getRecSpreadsheet.getRange('A6:C6').mergeAcross();

  if (percentValue > 75)
  { var thirdworldShort = "THIRD"
    var thirdworldLong = "Third World/Developing Country"
    var thirdworldReason
        = "Reason: Percent of inadequately managed plastic is "
          + percentValue + "% which is higher than 75%."

    //color&background formatting
        getRecRange1.setValue(thirdworldShort)
              .setFontStyle('italic')
              .setHorizontalAlignment('center')
              .setBackground('#cfe2f3');
        getRecRange2.setValue(thirdworldLong)
              .setHorizontalAlignment('left')
              .setBackground('#cfe2f3');
         getRecRange3.setValue(thirdworldReason)
               .setHorizontalAlignment('left')
               .setBackground('#cfe2f3');}

  else if (percentValue < 25)
  { var firstworldShort = "FIRST"
    var firstworldLong = "First World/Developed Country"
    var firstworldReason
        = "Reason: Percent of inadequately managed plastic is "
          + percentValue + "% which is lower than 25%."

    //color&background formatting
          getRecRange1.setValue(firstworldShort)
               .setFontStyle('italic')
               .setHorizontalAlignment('center')
               .setBackground('#cfe2f3');
          getRecRange2.setValue(firstworldLong)
               .setHorizontalAlignment('left')
               .setBackground('#cfe2f3');
          getRecRange3.setValue(firstworldReason)
               .setHorizontalAlignment('left')
               .setBackground('#cfe2f3');}

  else
  { var neither = "In the average range"
    var neitherReason
        = "Reason: Percent of inadequately managed plastic is "
          + percentValue + "% which is between 25% and 75%."

    //color&background formatting
           getRecRange1.setValue(neither)
                .setFontStyle('italic')
                .setHorizontalAlignment('center')
                .setBackground('#cfe2f3');
           getRecRange2.setValue(neitherReason)
                .setHorizontalAlignment('left')
                .setBackground('#cfe2f3');}
}
```

```
function activateGetRec() {

  var getRecSpreadsheet = SpreadsheetApp.getActiveSpreadsheet().getSheetByName('Get Recommendations UI');
  var userAns = getRecSpreadsheet.getCurrentCell().setFontWeight('bold').setFontColor('#3c78d8');
  userAns.copyTo(getRecSpreadsheet.getRange('E3'), SpreadsheetApp.CopyPasteType.PASTE_NORMAL, false);

  var Range = getRecSpreadsheet.getRange('A15:C15').activate();
  getRecSpreadsheet.getRange('A15:C15').mergeAcross();

  var Data = "Recommendations:"
  Range.setValue(Data).setFontWeight('bold').setHorizontalAlignment('left').setBackground('#9fc5e8');

  var userAnswerRange = getRecSpreadsheet.getRange('E3').setFontWeight('normal').setFontColor('black');
  var userAnswer = userAnswerRange.getValues();

  getRecommendations(userAnswer);}

function getRecommendations(userAnswer) {

  var Spreadsheet = SpreadsheetApp.getActiveSpreadsheet().getSheetByName('Get Recommendations UI');
  var countryStandingRange = Spreadsheet.getRange('A4');
  var countryStanding = countryStandingRange.getValues();
  var Spreadsheet2= SpreadsheetApp.getActiveSpreadsheet().getSheetByName('Recommendations Data Set');

  var count = 0;

  if (countryStanding == "FIRST") {
    var firstRange = Spreadsheet2.getRange('B2:B10');
    var firstValues = firstRange.getValues();
    var lastRow = firstValues.filter(String).length;

    var found = false;
    var foundRow;
    var targetRow = 15;

    for ( i = 0 ; i < lastRow ; i++){
      if(firstValues[i][0] == userAnswer){

        found = true;
        targetRow++;
        foundRow = i+2;

        if (found){
          var sourceRangeString='C'+foundRow;
          var sourceRange=Spreadsheet2.getRange(sourceRangeString);
          sourceRange.copyTo(Spreadsheet.getRange('A'+ targetRow ),
                               SpreadsheetApp.CopyPasteType.PASTE_NORMAL, false);
          count++;}
      }
    }
  }

  else if (countryStanding == "THIRD"){
    var thirdRange = Spreadsheet2.getRange('B11:B20');
    var thirdValues = thirdRange.getValues();
    var lastRow1 = thirdValues.filter(String).length;

    var found1 = false;
    var foundRow1;
    var targetRow1 = 15;

      for ( j = 0 ; j < lastRow1 ; j++){
        if(thirdValues[j][0] == userAnswer){

          found1 = true;
          foundRow1 = j+9;
          targetRow1++;

        if (found1){
          var sourceRangeString1='C'+foundRow1;
          var sourceRange1=Spreadsheet2.getRange(sourceRangeString1);
          sourceRange1.copyTo(Spreadsheet.getRange('A'+ targetRow1 ),
                               SpreadsheetApp.CopyPasteType.PASTE_NORMAL, false);
                count++;}
      }
    }
  }

  else {
    var midRange = Spreadsheet2.getRange('B2:B20');
    var midValues = midRange.getValues();
    var lastRow2 = midValues.filter(String).length;

    var found2 = false;
    var foundRow2;
    var targetRow2 = 15;

    for (k = 0; k < lastRow2 ; k++) {
      if(midValues[k][0] == userAnswer) {
        found2 = true;
        foundRow2 = k+2;
        targetRow2++;

        if (found2){
          var sourceRangeString2 = 'C' +foundRow2;
          var sourceRange2 = Spreadsheet2.getRange(sourceRangeString2);
          sourceRange2.copyTo(Spreadsheet.getRange('A' +targetRow2),
                               SpreadsheetApp.CopyPasteType.PASTE_NORMAL, false);
          count++;}
      }
    }
  }
  var range = Spreadsheet.getRange('E4');
  var values = range.getValue();
  range.setValue(count);
}
```

## Six Qualities of Code

The code for the program is high quality as it encompasses the six qualities of code and uses abstractions like the 2D array that maximizes the efficiency of the program. Also, the code incorporates for-loops and boolean values that help minimize the steps needed to get an output.

Decidable: In all cases of an input, there will be a corresponding output. In the method getCountryInfoByName(), the user can input any value (even if it is not a valid country name) and still get some type of output when the code is run. This way, the user has the option to choose to explore information about different countries. The following method determineCountryStanding() uses the percentValue found in the previous method. In the method askUserAbtRecommendaitons(), the user can also input any value other than YES or NO when they are asked if they want recommendations and still get an output. In the method getRecommendations(), the user can also choose any other cell in the spreadsheets other than HARD, MEDIUM, or EASY and the code will still get some output. The code is provable on paper. The code is complete because we can make a flowchart on paper as shown in Figure 6.

*Figure 6. Flow Chart*

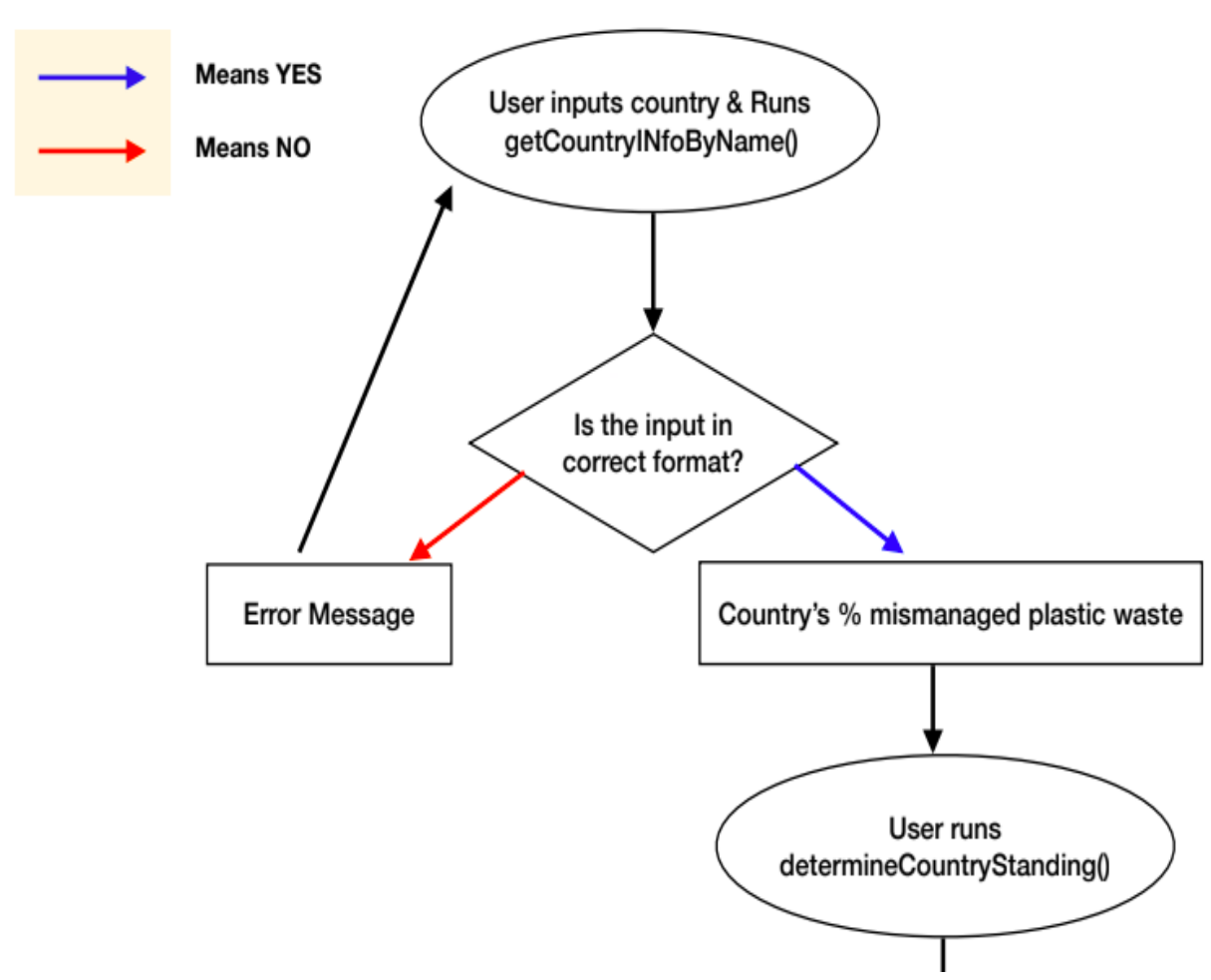

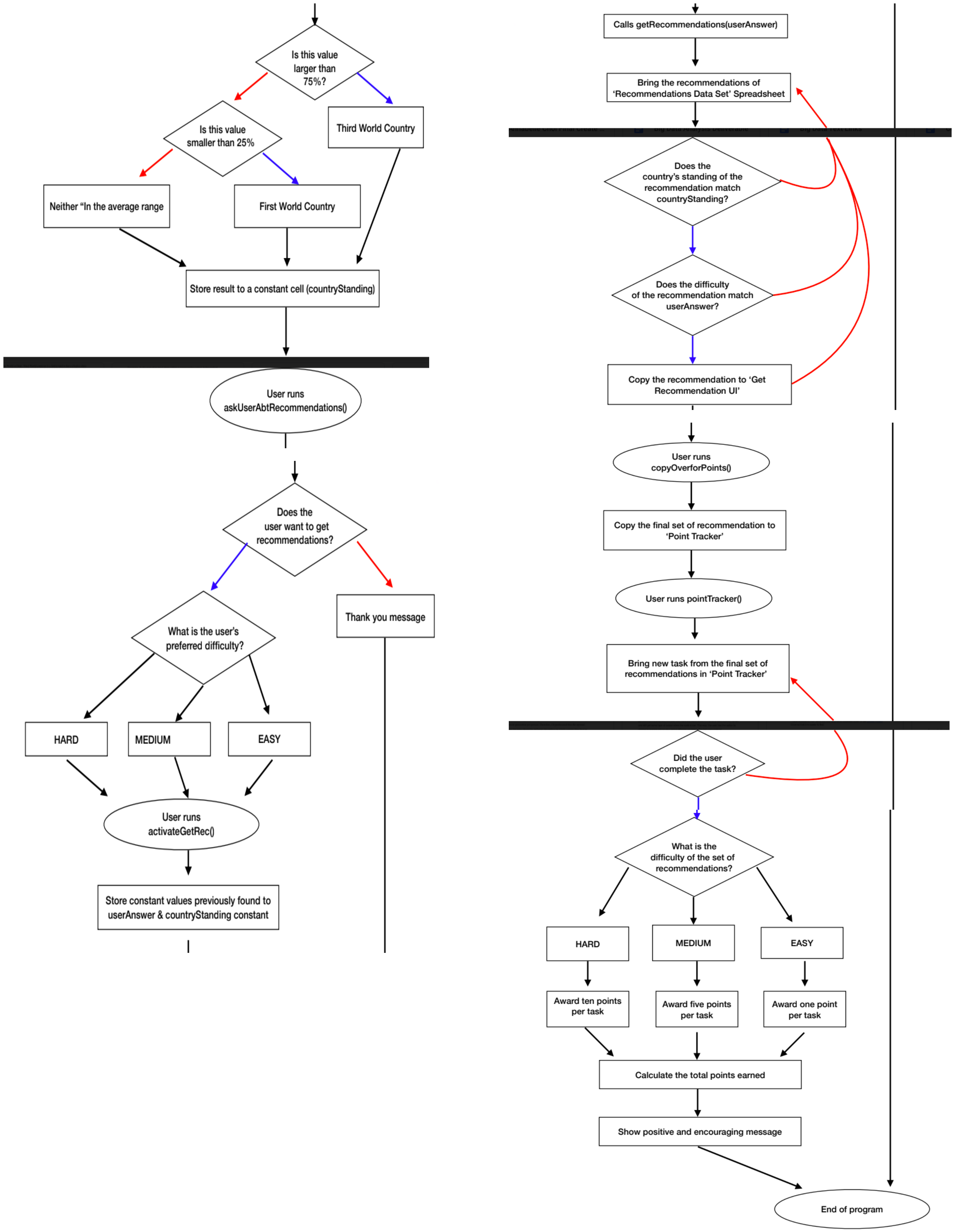

Is this value larger than 75%?
Third World Country
Is this value smaller than 25%
Neither "In the average range
First World Country
Store result to a constant cell (countryStanding)
User runs askUserAbtRecommendations()
Does the user want to get recommendations?
Thank you message
What is the user's preferred difficulty?
HARD
MEDIUM
EASY
User runs activateGetRec()
Store constant values previously found to userAnswer & countryStanding constant
Calls getRecommendations(userAnswer)
Bring the recommendations of 'Recommendations Data Set' Spreadsheet
Does the country's standing of the recommendation match countryStanding?
Does the difficulty of the recommendation match userAnswer?
Copy the recommendation to 'Get Recommendation UI'
User runs copyOverforPoints()
Copy the final set of recommendation to 'Point Tracker'
User runs pointTracker()
Bring new task from the final set of recommendations in 'Point Tracker'
Did the user complete the task?
What is the difficulty of the set of recommendations?
HARD
MEDIUM
EASY
Award ten points per task
Award five points per task
Award one point per task
Calculate the total points earned
Show positive and encouraging message
End of program

*Table 6. Truth Tables*

*Table 6-1. getCountryInfoByName() & determineCountryStanding()*

| Input | Output (% of inadequately managed plastic) from getCountryInfoByName | Country Standing Output from determineCountryStanding() |
|---|---|---|
| mexico | Country not found. Remember to type with first letter capital. | Reason: Percent of inadequately managed plastic is ()% |
| Mexico | 12% | FIRST, Reason: Percent of inadequately managed plastic is 12% which is lower than 25%. |
| bunny | Country not found. Remember to type with first letter capital. | Reason: Percent of inadequately managed plastic is ()% |
| Congo | 77% | THIRD, Reason: Percent of inadequately managed plastic is 77% which is higher than 75%. |
| Bulgaria | 31% | In the average range, Reason: Percent of inadequately managed plastic is 31% which is between 25% and 75%. |

*Table 6-2. askUserAbtRecommendations()*

| Input | Output |
|---|---|
| NO | Thank you for using our app! Come again soon :) |
| YES | How difficult would you like your recommendations to be? |
| yeet | Please reply with either YES or NO |
| no | Please reply with either YES or NO |

*Table 6-3. getRecommendations() & copyOverforPoints()*

| Input | Output of getRecommendations() | Output of copyOverforPoints() |
|---|---|---|
| C12 | 'EASY' copied in 'E3'<br>Will give corresponding recommendations<br><br>(ex) FIRST, EASY: Use a reusable straw instead of a plastic straw, Bring your own mug to a coffee shop, Post on social media about how your are part of the zero waste movement (ex) reddit sub, Start a social media trend advocating for zero waste such as hashtags | - Everything is pasted in Point Tracker Spreadsheet<br>- 4 in 'E4' |
| A12 | 'HARD' copied in 'E3'<br>Will give corresponding recommendations<br><br>(ex) FIRST, HARD: Bring awareness in your community about the national zero waste day and plan a project/activity the community can engage in, Make an impactful video advertisement promoting zero waste | - Everything is pasted in Point Tracker Spreadsheet<br>- 2 in 'E4' |
| A14 | Nothing copied in 'E3'<br>Will not give any recommendations | - Nothing copied to Point Tracker Spreadsheet<br>- 'E4' remains blank |
| A10 | 'YES" copied in 'E3'<br>Will not give any recommendations | - Nothing copied to Points Tracker Spreadsheet<br>- 'E4' remains blank |

Table 6-4. pointSystem()

| Input (difficulty, completed?) | Output |
|---|---|
| HARD, O | 10 |
| HARD, X | (0) |
| MEDIUM, O | 5 |
| EASY, O | 1 |
| EASY, X | (0) |

**Sound**

The code/program is logically entailed and provable. As shown in the truth tables, when we input true values, true values come out. For getCountryInfoByName() and determineCountryStanding, when we input a country then the program will determine the country's standing with proper reasoning. For askUserAbtRecommendations(), the inputs YES or NO will not give an error message and either terminate the code or move on to the next section of the code. For getRecommendations() & copyOverforPoints(), when we input the cells with HARD, MEDIUM, and EASY, the corresponding set of recommendations will show. Finally for pointSystem(), a certain point value shows if the user says that they have completed the task.

**Correct**

There seem to be no errors in our code when there is a correct input, the code will give information about the country, a set of recommendations, or the correct amount of total points (awarded for tasks that are accomplished) as we designed the algorithm.

**Efficient**

In our code, we optimized for efficiency. Our code is easy to follow and not cluttered with a lot of words. In each of the methods (except for copyOverforPoints()) uses an if-else block which ensures that code is organized into categories. For-loops and boolean values are used when choosing which recommendations meet the criteria which allow us to check for every single data point in the 'recommendation data set' array efficiently. I also used a 2D array in my last method to make code more efficient by helping manage the complexity of using two different variables at the same time. Using a 2D array allows me to store both difficulty level and user input. Thus, I can print out corresponding point values dynamically using a for loop, by simply changing the index numbers, making the final total calculation easier without creating many different variables.

**Polymorphic**

In our code, a variety of inputs can cause a method to behave differently which makes the code polymorphic, as shown in the truth tables. For example, depending on the user's country choice (affecting the country standing) and preferred difficulty level, the set of recommendations will be different. Also, depending on the preferred difficulty level of the recommendations and whether or not the user has accomplished those tasks, the number of total points awarded would be different.

## III. Conclusion & Suggestion for Future Research

For the current application, I have arbitrarily set the inequality values that determine if percent mismanaged plastic waste is high or low when determining if the country is first or third-world. For future modifications, I believe some background research would be needed to correctly determine what value and above percent mismanaged plastic should be considered third-world instead of estimating the value. I believe it would be worth adding more suggested recommendations especially for the countries that are in the average range because right now the code will randomly select recommendations made both for the first world countries and the third world countries. Also to statistically analyze exactly how efficient and motivating the application is for people to contribute to decreasing plastic pollution, I would like to do run an experiment in which users from both first-world and third-world countries use the application for a certain amount of time and rate their motivation levels to see if the application actually contributed to a significant change. The testing can be done simply through a google form and then we can use chi-squared analysis to see how significant the results were. The testing can also be done by calculating the change in total points earned to analyze in a more detailed fashion.

In terms of modifying the code, I suggest adding reset mechanisms on the code so that the spreadsheet automatically resets after running the entire application once. The tricky part though is that I will have to figure out a way to save the points values from the previous run into the next run so that the point values accumulate over time.

## References

[1] "Plastic Waste Generation per Person." *Our World in Data*, ourworldindata.org/grapher/plastic-waste-per-capita. Accessed 18 April 2020.

[2] "r/UrbanHell - People Offering Prayers at River Yamuna, India, which is Frothing from Industrial Waste." Reddit, https://www.reddit.com/r/UrbanHell/comments/9zmaj0/people_offering_prayers_at_river_yamuna_india/. Accessed 29 March 2020.

[3] "r/news - It's Raining Plastic: Microscopic Fibers Fall from the Sky in Rocky Mountains | US News." Reddit, https://www.reddit.com/r/news/comments/cpr

g1m/its_raining_plastic_microscopic_fibers_fall_from/. Accessed 29 March 2020.

[4] "Share of Plastic Waste That Is Inadequately Managed." *Our World in Data*, ourworldindata.org/grapher/inadequately-managed-plastic. Accessed 18 April 2020.